\documentclass[10pt,twocolumn, conference]{IEEEtran}

\usepackage{amsmath}
\usepackage{amssymb}
\usepackage[dvips]{graphicx}
\usepackage{epsfig}

\newcommand{\figsize}{0.38}

\begin{document}
\title{Achievable Rates and Training Optimization for Fading Relay Channels with Memory}
\author{\authorblockN{Sami Akin \hspace{0.4cm}
Mustafa Cenk Gursoy}
\authorblockA{Department of Electrical Engineering\\
University of Nebraska-Lincoln\\ Lincoln, NE 68588\\ Email:
sakin1@bigred.unl.edu, gursoy@engr.unl.edu}}
\date{}

\maketitle

\begin{abstract}
In this paper, transmission over time-selective, flat fading relay
channels is studied. It is assumed that channel fading coefficients
are not known a priori. Transmission takes place in two phases:
network training phase and data transmission phase. In the training
phase, pilot symbols are sent and the receivers employ single-pilot
MMSE estimation or noncausal Wiener filter to learn the channel.
Amplify-and-Forward (AF) and Decode-and-Forward (DF) techniques are
considered in the data transmission phase and achievable rate
expressions are obtained. The training period, and data and training
power allocations are jointly optimized by using the achievable rate
expressions. Numerical results are obtained considering Gauss-Markov
and lowpass fading models. Achievable rates are computed and
energy-per-bit requirements are investigated. The optimal power
distributions among pilot and data symbols are provided.
\end{abstract}

\section{Introduction}
In wireless communications, channel conditions vary randomly over
time due to mobility and changing environment. If the channel
conditions are not known a priori, practical wireless systems
generally employ pilot symbols to learn the channel. In one of the
early studies in this area, Cavers provided an analytical approach
to the design of pilot-assisted transmissions in \cite{Cavers1} and
\cite{Cavers2}. Considering adaptive coding of data symbols without
feedback to the transmitter, Abou-Faycal \emph{et al.}
\cite{faycal2} studied the data rates achieved with
pilot-symbol-assisted modulation (PSAM) over Gauss-Markov fading
channels. The authors in \cite{faycal3} also studied the PSAM over
Gauss-Markov channels and analyzed the power allocated to the data
symbols when the pilot symbol has fixed power. They showed that the
power of a data symbol decreases as its distance to the pilot symbol
increases.

Recently, cooperative wireless communications has attracted much
interest. Cooperative relay transmission techniques have been
studied in \cite{laneman} and \cite{laneman2} where
Amplify-and-Forward (AF) and Decode-and-Forward (DF) models are
considered. However, most of the studies have assumed that the
channel conditions are perfectly known at the receiver and/or
transmitter. In one of the recent studies, Wang \emph{et al.}
\cite{wang} considered wireless sensory relay networks where the
conditions of the channels are learnt imperfectly only by the relay
nodes.

In this paper, we study the training-based transmission and
reception schemes over a priori unknown, Rayleigh fading relay
channels in which the fading is modeled as a random process with
memory. Unknown fading coefficients of the channels are estimated at
the receivers with the assistance of the pilot symbols. We consider
two channel estimation methods: single-pilot minimum
mean-square-error (MMSE) estimation and noncausal Wiener filter
estimation. We study AF and DF relaying techniques with two
different transmission protocols. We obtain achievable rate
expressions and optimize the training parameters by maximizing these
expressions. We concentrate on the Gauss-Markov and lowpass fading
processes for numerical analysis.

\section{Channel Model}
We consider a three-node-relay network which consists of one source
node, one relay node and one destination node. Source-destination,
source-relay and relay-destination channels are modeled as Rayleigh
fading channels with fading coefficients denoted by
$h_{sd}\sim\mathcal{CN}(0,\sigma_{sd}^{2})$,
$h_{sr}\sim\mathcal{CN}(0,\sigma_{sr}^{2})$ and
$h_{rd}\sim\mathcal{CN}(0,\sigma_{rd}^{2})$\footnote{$x \sim
\mathcal{CN}(m,\sigma^{2})$ is used to denote that $x$ is a proper
complex Gaussian random variable with mean $m$ and variance
$\sigma^2$.}, respectively. Each channel is independent of others
and exhibits memory with an arbitrary correlation structure. Hence
$\{h_{sd}\}$, $\{h_{sr}\}$, and $\{h_{rd}\}$ are assumed to be
mutually independent Gaussian random processes with power spectral
densities $S_{h_{sd}}(e^{jw})$, $S_{h_{sr}}(e^{jw})$ and
$S_{h_{rd}}(e^{jw})$, respectively. In this relay network,
information is sent from the source to the destination with the aid
of the relay. Transmission takes place in two phases: network
training phase and data transmission phase. Over a duration of $M$
symbols, the source and the relay are subject to the following power
constraints:
\begin{gather}\label{power-constraint}
\|\textbf{x}_{s,t}\|^{2}+E\{\|\textbf{x}_{s}\|^{2}\}\leq M P_s
\\
\|\textbf{x}_{r,t}\|^{2}++E\{\|\textbf{x}_{r}\|^{2}\} \le M P_r
\end{gather}
where $\textbf{x}_{s,t}$ and $\textbf{x}_{r,t}$ are the source and
relay pilot vectors, respectively, and $\textbf{x}_{s}$ and
$\textbf{x}_{r}$ are the data vectors sent by the source and the
relay, respectively.

\section{Network Training Phase}
In the network training phase, source and relay send pilot symbols
in nonoverlapping intervals with a period of $M$ symbols to
facilitate channel estimation at the receivers. In a block of $M$
symbols, transmission takes place in the following order. First, the
source sends a single pilot symbol $x_{s,t}$, and the relay and
destination receives
\begin{align}\label{received-signals1}
y_{r,t}=h_{sr}x_{s,t}+n_{r} \quad \text{and} \quad
y_{d,t}=h_{sd}x_{s,t}+n_{d},
\end{align}
and estimates $h_{sr}$ and $h_{sd}$, respectively. Then,
transmission enters the data transmission phase, and source sends an
$(M-2)/2$-dimensional data vector that is again received by the
relay and destination terminals. Next, only the relay sends a single
pilot symbol $x_{r,t}$, and the signal received at the destination
node is
\begin{equation}\label{received-signals2}
y_{d,t}^{r}=h_{rd}x_{r,t}+n_{d}^r,
\end{equation}
which is used by the destination to estimate $h_{rd}$. In
(\ref{received-signals1}) and (\ref{received-signals2}), $n_{r}$,
$n_{d}$, and  $n_d^r$ are assumed to be independent and identically
distributed (i.i.d.) zero mean Gaussian random variables with
variance $\sigma_{n}^{2}$, modeling the additive thermal noise
present at the receivers. In the remaining duration of $(M-2)/2$
symbols, transmission again enters the data transmission phase. In
this case, the relay transmits an $(M-2)/2$ dimensional data vector
to the destination while the source either becomes silent or
continues its transmission depending on the cooperation protocol.
This order of transmission is repeated for the next block of $M$
symbols.

As noted before, we consider two channel estimation methods. In the
first method, only a single pilot symbol is used to obtain the MMSE
estimate of the channel fading coefficients.
%
As described in $\cite{Gursoy1}$, MMSE estimates of the fading
coefficients and the variances of the estimate errors are given as
follows\footnote{$\widehat{h}$ and $\widetilde{h}$ are used to
denote the estimate and error in the estimate of $h$, respectively.
Hence, we can write $h = \widehat{h}+\widetilde{h}$.}:
\begin{equation}\label{hsr}
\widehat{h}_{sr}=\frac{\sigma_{sr}^{2}\sqrt{P_{x_{s,t}}}}{\sigma_{sr}^{2}P_{x_{s,t}}+\sigma_{n}^{2}}y_{r,t},
\quad
\sigma_{\widetilde{h}_{sr}}^{2}=\frac{\sigma_{sr}^{2}\sigma_{n}^{2}}{\sigma_{sr}^{2}P_{x_{s,t}}+\sigma_{n}^{2}}
\end{equation}
\begin{equation}\label{hsd}
\widehat{h}_{sd}=\frac{\sigma_{sd}^{2}\sqrt{P_{x_{s,t}}}}{\sigma_{sd}^{2}P_{x_{s,t}}+\sigma_{n}^{2}}y_{d,t},
\quad
\sigma_{\widetilde{h}_{sd}}^{2}=\frac{\sigma_{sd}^{2}\sigma_{n}^{2}}{\sigma_{sd}^{2}P_{x_{s,t}}+\sigma_{n}^{2}}
\end{equation}
\begin{equation}\label{hrd}
\widehat{h}_{rd}=\frac{\sigma_{rd}^{2}\sqrt{P_{x_{r,t}}}}{\sigma_{rd}^{2}P_{x_{r,t}}+\sigma_{n}^{2}}y_{d,t}^{r},
\quad
\sigma_{\widetilde{h}_{rd}}^{2}=\frac{\sigma_{rd}^{2}\,\sigma_{n}^{2}}{\sigma_{rd}^{2}P_{x_{s,t}}+\sigma_{n}^{2}}
\end{equation}
where $P_{x_{s,t}}$ and $P_{x_{r,t}}$ are the power of the pilot
symbols sent by the source and the relay, respectively, and
$y_{r,t}\sim\mathcal{CN}(0,\sigma_{sr}^{2}P_{x_{s,t}}+\sigma_{n}^{2})$,
$y_{d,t}\sim\mathcal{CN}(0,\sigma_{sd}^{2}P_{x_{s,t}}+\sigma_{n}^{2})$
and
$y_{d,t}^{r}\sim\mathcal{CN}(0,\sigma_{rd}^{2}P_{x_{r,t}}+\sigma_{n}^{2})$.

In the second method, we employ the noncausal Wiener filter which is
the optimum linear estimator in the mean-square sense. The Wiener
filter is employed at both the relay and the destination. Note that
since pilot symbols are sent with a period of $M$ symbols, the
channels are sampled every $MT_{s}$ seconds, where $T_{s}$ is the
sampling time. As described in $\cite{Sam-Gur}$, we have to consider
the undersampled versions of the Doppler spectrums of the fading
coefficients, which are given by
\begin{equation}\label{undersampled-sr}
S_{h_{sr},m}(e^{jw})=\frac{1}{M}\sum_{k=0}^{M-1}e^{jm(w-2\pi
k)/M}S_{h_{sr}}\left(e^{j(w-2\pi k)/M}\right)
\end{equation}
\begin{equation}\label{undersampled-sd}
S_{h_{sd},m}(e^{jw})=\frac{1}{M}\sum_{k=0}^{M-1}e^{jm(w-2\pi
k)/M}S_{h_{sd}}\left(e^{j(w-2\pi k)/M}\right)
\end{equation}
\begin{equation}\label{undersampled-rd}
S_{h_{rd},m}(e^{jw})=\frac{1}{M}\sum_{k=0}^{M-1}e^{jm(w-2\pi
k)/M}S_{h_{rd}}\left(e^{j(w-2\pi k)/M}\right).
\end{equation}
Then, the channel MMSE variances for the noncausal Wiener filter at
time $Ml+m$ are given by \cite{book}
\begin{equation}\label{estimation error-sr}
\sigma_{\widetilde{h}_{sr}}^{2}(Ml+m)=\sigma_{sr}^{2}-\frac{1}
{2\pi}\int_{-\pi}^{\pi}\frac{P_{x_{s,t}}|S_{h_{sr},m}(e^{jw})|^{2}}
{P_{x_{s,t}}S_{h_{sr},0}(e^{jw})+\sigma_{n}^{2}}dw
\end{equation}
\begin{equation}\label{estimation error-sd}
\sigma_{\widetilde{h}_{sd}}^{2}(Ml+m)=\sigma_{sd}^{2}-\frac{1}
{2\pi}\int_{-\pi}^{\pi}\frac{P_{x_{s,t}}|S_{h_{sd},m}(e^{jw})|^{2}}
{P_{x_{s,t}}S_{h_{sd},0}(e^{jw})+\sigma_{n}^{2}}dw
\end{equation}
\begin{equation}\label{estimation error-rd}
\sigma_{\widetilde{h}_{rd}}^{2}(Ml+m)=\sigma_{rd}^{2}-\frac{1}
{2\pi}\int_{-\pi}^{\pi}\frac{P_{x_{r,t}}|S_{h_{rd},m}(e^{jw})|^{2}}
{P_{x_{r,t}}S_{h_{rd},0}(e^{jw})+\sigma_{n}^{2}}dw
\end{equation}
\begin{gather*}
\text{for } \quad l=0,1,2, \ldots \quad \text{and} \quad m=0,1\dots
(M-1).
\end{gather*}
After obtaining the estimates, we can express the fading
coefficients as
\begin{align}\label{newfadingcoeffs}
&h_{sr}(Ml+m)=\widehat{h}_{sr}(Ml+m) + \widetilde{h}_{sr}(Ml+m)\nonumber \\
&h_{sd}(Ml+m)=\widehat{h}_{sd}(Ml+m) + \widetilde{h}_{sd}(Ml+m)\nonumber \\
&h_{rd}(Ml+m)=\widehat{h}_{rd}(Ml+m) + \widetilde{h}_{rd}(Ml+m).
\end{align}
\section{Data Transmission Phase}

Note that as described in the previous section, within a block of
$M$ symbols, two symbol durations are allocated for channel training
while data transmission is performed in the remaining portion of the
time. We assume that relay operates in half-duplex mode. Hence, the
relay first listens and
then transmits to the destination. 
We consider two transmission protocols: non-overlapped and
overlapped transmissions.

\subsection{Non-overlapped Case}
In this protocol, the source and relay send data symbols in
nonoverlapping intervals.  The source, after sending the pilot
symbol, sends its $(M-2)/2$ data symbols which are received by the
relay and the destination as\footnote{Since we consider transmission
in a block of $M$ symbols, we drop the block index for the sake of
simplicity and use $m$ instead of using $Ml+m$.}
\begin{align}\label{received-signals11}
&y_{r,d}(m)=h_{sr}(m)x_{s,d}(m)+n_{r}(m)\\ \nonumber
&y_{d,d}(m)=h_{sd}(m)x_{s,d}(m)+n_{d}(m) \quad m=2,\ldots
\frac{M}{2}.
\end{align}
Next, the source stops transmission, and the relay sends first its
pilot symbol and then  $(M-2)/2$ data symbols which are generated
from $\textbf{y}_{r,d} = [y_{r,d}(2), \ldots, y_{r,d}(M/2)]$. Thus
the destination receives
\begin{equation}\label{received-signals12}
y_{d,d}(j)=h_{rd}(j)x_{r,d}(j)+n_{d}(j)\quad j=m+M/2
\end{equation}
where again $m = 2,\ldots, M/2$. After substituting
(\ref{newfadingcoeffs}) into (\ref{received-signals11}) and
(\ref{received-signals12}), we obtain
\begin{align}\label{ensonki output}
&y_{r,d}(m)=\widehat{h}_{sr}(m)x_{s,d}(m)+\widetilde{h}_{sr}(m)x_{s,d}(m)+n_{r}(m)
\\ \nonumber
&y_{d,d}(m)=\widehat{h}_{sd}(m)x_{s,d}(m)+\widetilde{h}_{sd}(m)x_{s,d}(m)+n_{d}(m)
\\ \nonumber
&y_{d,d}(j)=\widehat{h}_{rd}(j)x_{r,d}(j)+\widetilde{h}_{rd}(j)x_{r,d}(j)+n_{d}(j)
\end{align}
where $m = 2,\ldots, M/2$ and $j=m+M/2$.
\subsection{Overlapped Case}

In this protocol, the source continues its transmission while the
relay is sending its data symbols. The source becomes silent only
when the relay is sending the pilot symbol. Therefore, the received
signals in the data transmission phase can be written as
\begin{align}\label{received-signals overlapped}
&y_{r,d}(m)=h_{sr}(m)x_{s,d}(m)+n_{r}(m)\\ \nonumber
&y_{d,d}(m)=h_{sd}(m)x_{s,d}(m)+n_{d}(m)\\ \nonumber
&y_{d,d}(j)=h_{sd}(j)x_{s,d}(j)+h_{rd}(j)x_{r,d}(j)+n_{d}(j)
\end{align} where $m=2,...,M/2$ and $j=m+M/2$.
Similarly as in the non-overlapped case, we can integrate the
estimation results to (\ref{received-signals overlapped}) and write
\begin{align}\label{ensonki output for overlapped case}
&y_{r,d}(m)=\widehat{h}_{sr}(m)x_{s,d}(m)+\widetilde{h}_{sr}(m)x_{s,d}(m)+n_{r}(m)
\\ \nonumber
&y_{d,d}(m)=\widehat{h}_{sd}(m)x_{s,d}(m)+\widetilde{h}_{sd}(m)x_{s,d}(m)+n_{d}(m)
\\ \nonumber
&y_{d,d}(j)=\widehat{h}_{sd}(j)x_{s,d}(j)+\widetilde{h}_{sd}(j)x_{s,d}(j)\\
\nonumber
&+\widehat{h}_{rd}(j)x_{r,d}(j)+\widetilde{h}_{rd}(j)x_{r,d}(j)+n_{d}(j).
\end{align}

\section{Achievable Rates for AF Scheme}

In this section, we consider the AF relaying scheme in which the
relay sends to the destination simply the scaled version of the
signal received from the source. An achievable rate expression for
the AF scheme is obtained by maximizing the mutual information
between the transmitted signal vector $\textbf{x}_{s,d}$ and the
$(M\!\!\!-\!\!\!2)$-dimensional received signal $\textbf{y}_{d,d} =
[y_{d,d}(2),\ldots,y_{d,d}(M/2),y_{d,d}(M/2+2),\ldots,y_{d,d}(M)]$
given the estimates of the fading coefficients.
$\widehat{\textbf{h}}_{sr}$, $\widehat{\textbf{h}}_{sd}$ and
$\widehat{\textbf{h}}_{rd}$ are used to denote the vectors of
channel estimates. Therefore, an achievable rate expression is given
by
\begin{equation}\label{capacity}
I_{AF}=\sup_{p_{\textbf{x}_{s}}}\frac{1}{M}\textit{I}(\textbf{x}_{s,d};\textbf{y}_{d,d}|\widehat{\textbf{h}}_{sr},\widehat{\textbf{h}}_{sd},
\widehat{\textbf{h}}_{rd}).
\end{equation}
Note that the above formulation supposes that the destination node
also knows $\widehat{\textbf{h}}_{sr}$. Hence, it is assumed that
these estimates are reliably forwarded by the relay to the
destination using low rate links. A lower bound on $I_{AF}$ can be
obtained by assuming similarly as in $\cite{Hassibi}$ that the
estimation errors are additional sources of worst-case Gaussian
noise. We define the new noise random variables in non-overlapped
and overlapped cases as
\begin{align}\label{error 1}
&z_{r,d}(m)=\widetilde{h}_{sr}(m)x_{s,d}(m)+n_{r}(m) \\ \nonumber
&z_{d,d}(m)=\widetilde{h}_{sd}(m)x_{s,d}(m)+n_{d}(m) \\ \nonumber
&z_{d,d}(j)=\widetilde{h}_{rd}(j)x_{r,d}(j)+n_{d}(j)
\end{align}
and
\begin{align}\label{error 2}
&z_{r,d}(m)=\widetilde{h}_{sr}(m)x_{s,d}(m)+n_{r}(m) \\ \nonumber
&z_{d,d}(m)=\widetilde{h}_{sd}(m)x_{s,d}(m)+n_{d}(m) \\ \nonumber
&z_{d,d}(j)=\widetilde{h}_{sd}(j)x_{s,d}(j)+\widetilde{h}_{rd}(j)x_{r,d}(j)+n_{d}(j)
\end{align}
respectively. By assuming that the new noise components are Gaussian
random variables and
%
%
using techniques similar to those in $\cite{jun-gur}$, we can obtain
the following worst-case achievable rate expression for the
non-overlapped case:
\begin{equation}\label{cworst-non-overlapped}
I_{nonover}=\frac{1}{M}E_{w_{sr}}E_{w_{sd}}E_{w_{rd}}\sum_{m=2}^{M/2}
\log \left(1+a_{1,m}+f(b_{1,m},c_{1,j})\right)
\end{equation}
where
\begin{equation}\label{a1b1c1}
a_{1,m}=\frac{P_{x_{s,d}(m)}\sigma_{\widehat{h}_{sd}(m)}^{2}}{\sigma_{z_{d,d}(m)}^{2}}|w_{sd}|^{2},
\quad
b_{1,m}=\frac{P_{x_{s,d}(m)}\sigma_{\widehat{h}_{sr}(m)}^{2}}{\sigma_{z_{r,d}(m)}^{2}}|w_{sr}|^{2},
\end{equation}
%
\begin{equation*}
c_{1,j}=\frac{P_{x_{r,d}(j)}\sigma_{\widehat{h}_{rd}(j)}^{2}}{\sigma_{z_{d,d}(j)}^{2}}|w_{rd}|^{2},\quad
f(x,y)=\frac{xy}{1+x+y}
\end{equation*}
%
and $w_{sd}\sim\mathcal{CN}(0,1)$, $w_{sr}\sim\mathcal{CN}(0,1)$,
$w_{rd}\sim\mathcal{CN}(0,1)$. $P_{x_{s,d}(m)}$ and $P_{x_{r,d}(j)}$
are the powers of the $m^{th}$ source symbol and $j^{th}$ relay
symbol, respectively, and
$\sigma_{\widehat{h}_{sr}(m)}^{2}=\sigma_{sr}^{2}-\sigma_{\widetilde{h}_{sr}(m)}^{2}$,
$\sigma_{\widehat{h}_{sd}(m)}^{2}=\sigma_{sd}^{2}-\sigma_{\widetilde{h}_{sd}(m)}^{2}$,
$\sigma_{\widehat{h}_{rd}(m)}^{2}=\sigma_{rd}^{2}-\sigma_{\widetilde{h}_{rd}(m)}^{2}$.
Finally, note that $j=m+M/2$.

Similarly, we can find the following achievable rate expression for
the
overlapped case:
\begin{align}\label{cworst-overlapped}
I_{over}&=\frac{1}{M}E_{w_{sr}}E_{w_{sd}}E_{w_{rd}}\\
\nonumber
&\sum_{m=2}^{M/2}\log\left(1+a_{2,m}+f(d_{2,m},c_{2,j})+q(a_{2,m},b_{2,j},c_{2,j},d_{2,m})\right)
\end{align}
where
\begin{equation}\label{a2b2c2d2}
a_{2,m}=\frac{P_{x_{s,d}(m)}\sigma_{\widehat{h}_{sd}(m)}^{2}}{\sigma_{z_{d,d}(m)}^{2}}|w_{sd}|^{2},
b_{2,j}=\frac{P_{x_{s,d}(j)}\sigma_{\widehat{h}_{sd}(j)}^{2}}{\sigma_{z_{d,d}(j)}^{2}}|w_{sd}|^{2},
\end{equation}
\begin{equation*}
c_{2,j}=\frac{P_{x_{r,d}(j)}\sigma_{\widehat{h}_{rd}(j)}^{2}}{\sigma_{z_{d,d}(j)}^{2}}|w_{rd}|^{2},
d_{2,m}=\frac{P_{x_{s,d}(m)}\sigma_{\widehat{h}_{sr}(m)}^{2}}{\sigma_{z_{r,d}(m)}^{2}}|w_{sr}|^{2}
\end{equation*}
and
$
q(a,b,c,d)=\frac{(1+a)b(1+c)}{1+c+d} \quad \text{and} \quad j = m +
M/2.
$

\section{Achievable Rates for DF Scheme}
The repetition coding and the parallel coding are two possible
coding techniques used in DF schemes \cite{laneman}. First, we
consider the repetition coding, and for this case the achievable
rate is given by
\begin{equation}\label{repetition capacity}
I_{rc}=\frac{1}{M}\sup_{p_{\textbf{x}_{s}}}\left\{
\min\left[\textit{I}(\textbf{x}_{s,d};\textbf{y}_{r,d}|\widehat{\textbf{h}}_{sr}),
\textit{I}(\textbf{x}_{s,d};\textbf{y}_{d,d}|\widehat{\textbf{h}}_{sd},\widehat{\textbf{h}}_{rd})
\right]\right\}
\end{equation}
Employing the techniques used in the AF non-overlapped scheme, we
obtain the following achievable rate expression for non-overlapped
DF with repetition coding:
\begin{equation}\label{cworst df repet non over}
I_{nonover,rc}=\frac{1}{M}E_{w_{sr}}E_{w_{sd}}E_{w_{rd}}\sum_m
\min(I_{1},I_{2})
\end{equation}
where
\begin{equation*}\label{I1}
I_{1}=\log\left[1+b_{1,m}\right],\quad
I_{2}=\log\left[1+a_{1,m}+c_{1,j} \right],
\end{equation*}
and $a_{1,m},b_{1,m}$ and $c_{1,j}$ are given in (\ref{a1b1c1}).
For the overlapped case of the DF repetition coding, (\ref{cworst df
repet non over}) holds with  $I_{1}$ and $I_{2}$ defined as
\begin{equation*}\label{I1}
I_{1}=\log\left[1+c_{2,j}\right], \quad
I_{2}=\log\left[1+a_{2,m}+b_{2,j}+d_{2,m}+a_{2,m}b_{2,j}\right]
\end{equation*}
where $a_{2,m},b_{2,j},c_{2,j},$ and $d_{2,m}$ are given in
(\ref{a2b2c2d2}).
When we employ the parallel coding, we have
\begin{align}\label{capacity parallel}
&I_{pc}=\frac{1}{M}\sup_{P_{\textbf{x}_{s}},P_{\textbf{x}_{r}}}\\
\nonumber
&\left\{\min\left[\textit{I}(\textbf{x}_{s,d};\textbf{y}_{r,d}|\widehat{\textbf{h}}_{sr}),
\textit{I}(\textbf{x}_{s,d};\textbf{y}_{d,d}|\widehat{\textbf{h}}_{sd})
+\textit{I}(\textbf{x}_{r,d};\textbf{y}_{d,d}|\widehat{\textbf{h}}_{rd})
\right]\right\}.
\end{align}
Similarly, we can find, for the nonoverlapping case, an achievable
rate expression given by
\begin{equation}\label{cworst df para non over}
I_{nonover,pc}=\frac{1}{M}E_{w_{sr}}E_{w_{sd}}E_{w_{rd}}\sum_{m}\min(I_{1},I_{2})
\end{equation}
where
\begin{equation*}\label{I1}
I_{1}=\log(1+b_{1,m}),\quad I_{2}=\log(1+a_{1,m})+\log(1+c_{1,j}).
\end{equation*}

\begin{figure}
\begin{center}
\includegraphics[width = \figsize\textwidth]{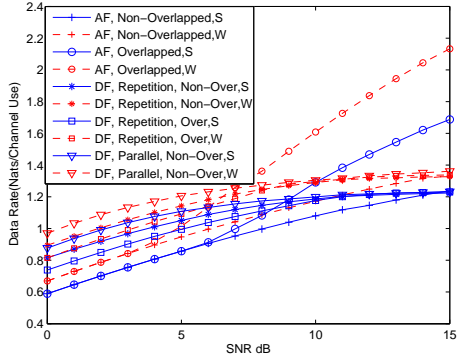}
\caption{The optimal achievable rates vs. SNR for the Gauss-Markov
fading model ($\alpha = 0.99$) and different relaying techniques.
$\sigma_{sd}^{2}=1,\sigma_{sr}^{2}=16$ and $\sigma_{rd}^{2}=16$. (S:
single-pilot estimation. W: Wiener filter.)} \label{fig:fig1}
\end{center}
\end{figure}
\begin{figure}
\begin{center}
\includegraphics[width = \figsize\textwidth]{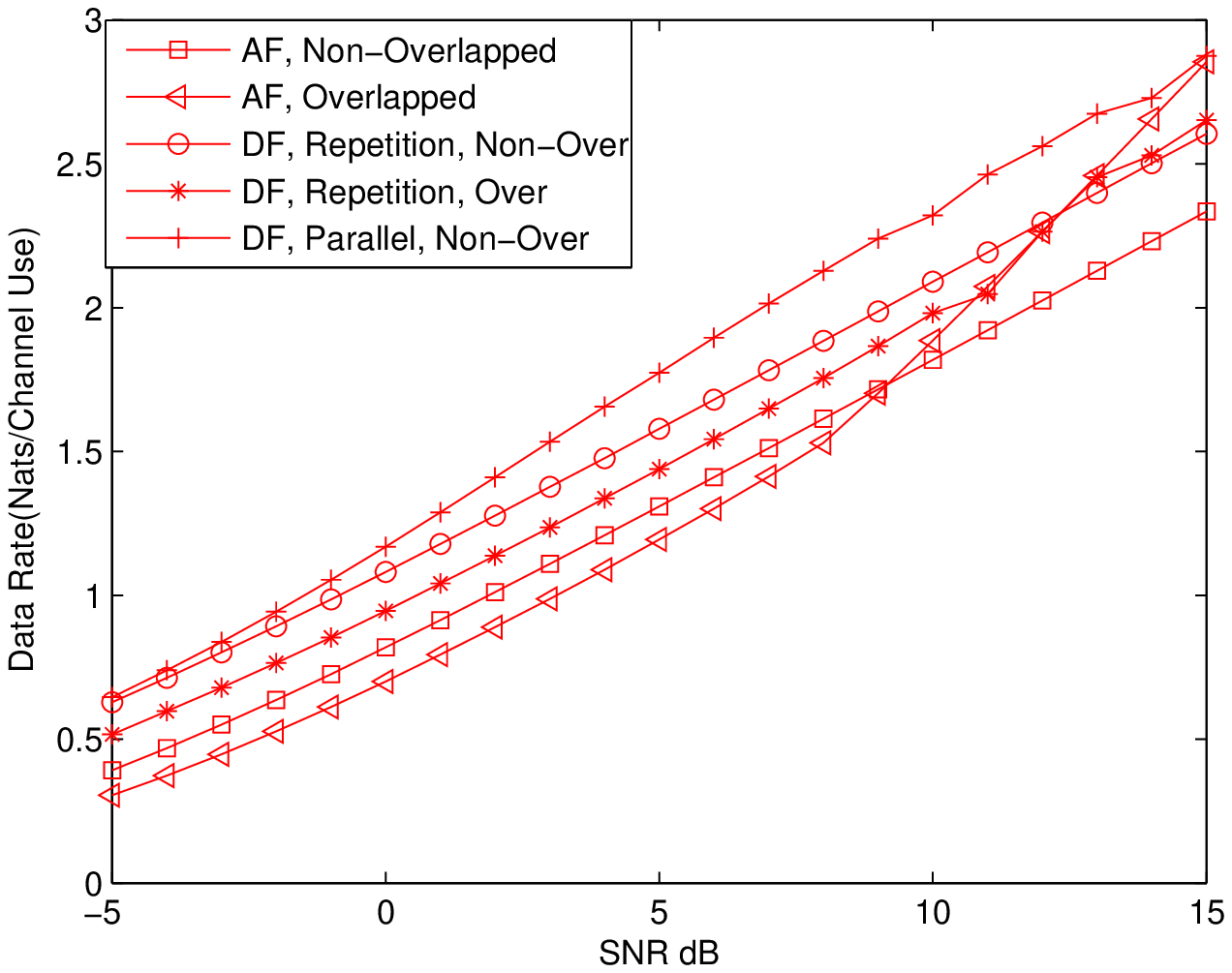}
\caption{The optimal achievable rates vs. SNR for the lowpass fading
model when noncausal Wiener filter is employed.
$\sigma_{sd}^{2}=1,\sigma_{sr}^{2}=16$ and $\sigma_{rd}^{2}=16$.}
\label{fig:fig2}
\end{center}
\end{figure}
\begin{figure}
\begin{center}
\includegraphics[width = \figsize\textwidth]{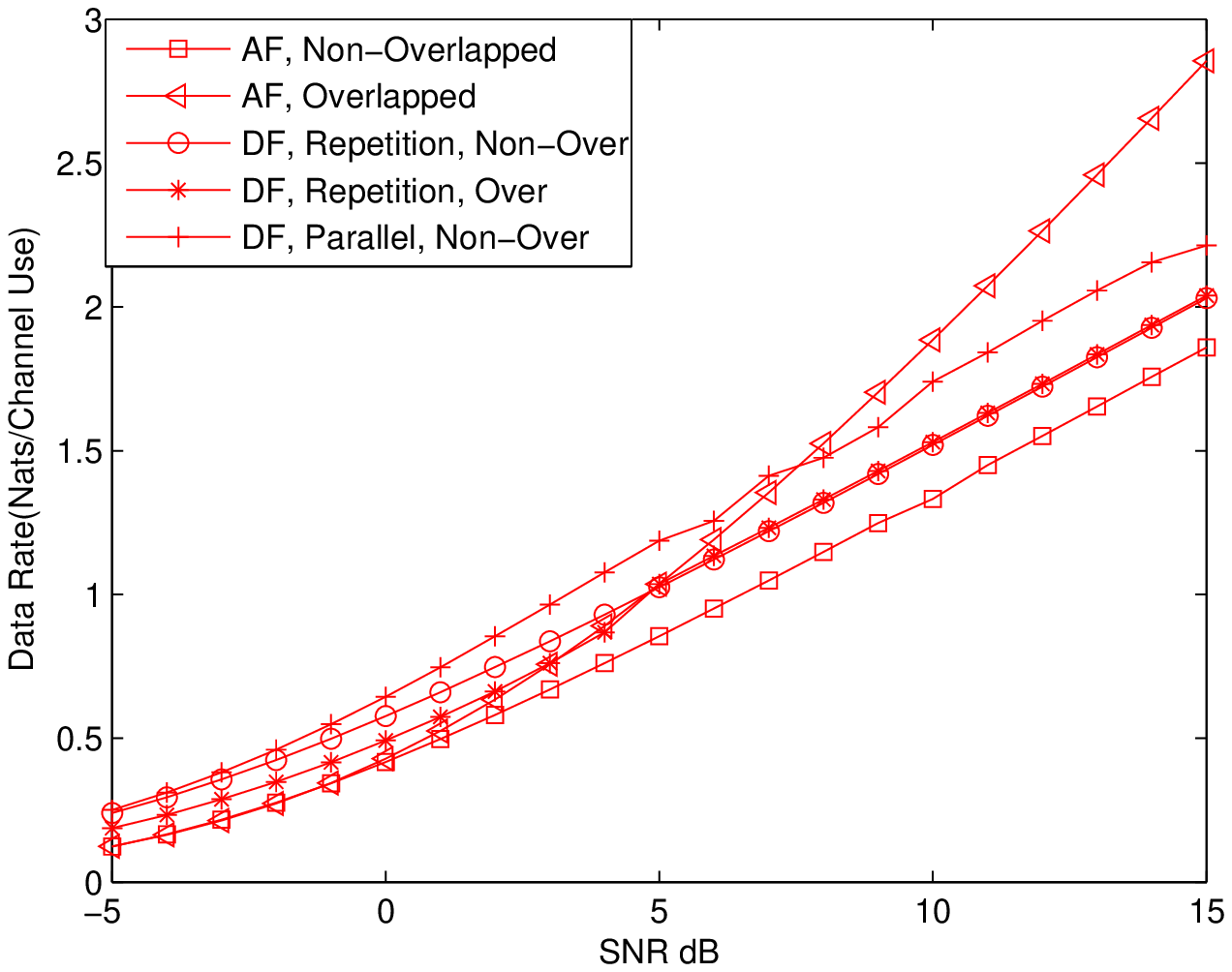}
\caption{The optimal achievable rates vs. SNR for the lowpass fading
model when noncausal Wiener filter is employed.
$\sigma_{sd}^{2}=1,\sigma_{sr}^{2}=4$ and $\sigma_{rd}^{2}=4$.}
\label{fig:fig3}
\end{center}
\end{figure}
\begin{figure}
\begin{center}
\includegraphics[width = \figsize\textwidth]{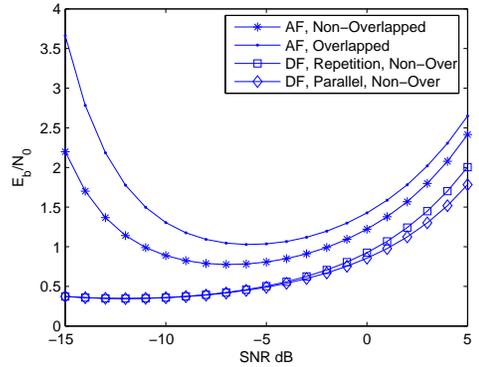}
\caption{Normalized bit energies $\frac{E_{b}}{N_{0}}$ vs. SNR for
the lowpass fading model when noncausal Wiener filter is employed.
$\sigma_{sd}^{2}=1,\sigma_{sr}^{2}=16$ and $\sigma_{rd}^{2}=16$.}
\label{fig:fig4}
\end{center}
\end{figure}

\section{Optimizing Training Parameters}
In this section, we consider two particular fading processes. In the
first case, fading is modeled as a first-order Gauss-Markov process
whose dynamics is described by
\begin{equation*}
h_{k}=\alpha h_{k-1}+z_{k} \quad 0\leq\alpha\leq1\quad
k=1,2,3,\ldots
\end{equation*}
where \{$z_{k}$\} are i.i.d. circular complex Gaussian variables
with zero mean and variance equal to (1-$\alpha^{2}) \sigma_{h}^2$.
In the above formulation, $\alpha$ is a parameter that controls the
rate of the channel variations between consecutive transmissions.
The power spectral density of the Gauss-Markov process with variance
$\sigma_{h}^{2}$ is given by
\begin{equation}\label{psd}
S_h(e^{jw})=\frac{(1-\alpha^{2})\sigma_{h}^{2}}{1+\alpha^{2}-2\alpha\cos(w)}.
\end{equation}
We also model the fading as a lowpass Gaussian process whose power
spectral density is given by
\begin{equation}\label{psd_low}
S_h(e^{jw})=\left\{\begin{split}
              &\frac{\sigma_{h}^{2}}{2f_{d}},\quad &\text{for } |w|<w_{d}\\
              &0,\quad &\text{otherwise}
            \end{split}\right.
\end{equation}
where $w_d = 2 \pi f_d$ is the maximum Doppler spread in radians.

In Gauss-Markov channels, it is difficult to find a closed-form
expression for the variance of the estimate error when Wiener filter
is used, because the channel's spectrum is not band limited.
Therefore, there is always aliasing in the undersampled Doppler
spectrums, which causes an increase in the variance of the error. On
the other hand, when fading is modeled as a lowpass process, we can
find a explicit solution for the error variance, and we can express
it as
\begin{equation*}
\sigma_{\widetilde{h}}^{2}=\frac{\sigma_{h}^{2}\sigma_{n}^{2}}{P_{x,t}\sigma_{h}^{2}+\sigma_{n}^{2}}.
\end{equation*}
In the lowpass case, if the channel is sampled sufficiently fast
(i.e., $M < \frac{1}{2f_d}$), there is no aliasing and the power is
distributed equally among data symbols. However, note that the power
allocated to the data symbols of the source is not equal to the
power allocated to the data symbols of the relay. In general, if
there is aliasing or a single pilot is used for estimation, the
power allocated to the data symbols will differ depending on their
distance to the pilot signals.

Having obtained achievable rate expressions, our next goal is to
jointly optimize training period $M$, training power, and power
allocated to the data symbols.


\section{Numerical Results}

In this section, we present numerical optimization results. 
In Figure \ref{fig:fig1}, we plot the optimal achievable rates with
respect to SNR for different relaying protocols by using two
different methods of channel estimation. Fading is assumed to be a
Gauss-Markov process. The dashed lines indicate the optimal data
rates obtained when noncausal Wiener filter is used, whereas the
solid lines show the optimal data rates obtained when a single pilot
symbol is used for estimation. The rates are optimal in the sense
that they are obtained with optimal training parameters and optimal
power allocations. We can see that at low SNR values, DF provides
higher rates and parallel non-overlapped DF scheme is the most
efficient one. As expected, Wiener filter performance is better than
that of the estimation that uses a single pilot. Moreover, at low
SNR values non-overlapped and overlapped relaying schemes give the
same optimal results, and optimal power distributions among data and
pilot symbols are the same for both. On the other hand, at high SNR
values, we see a significant increase in the data rate of AF
overlapped scheme compared to the other schemes.

\begin{figure}
\begin{center}
\includegraphics[width = \figsize\textwidth]{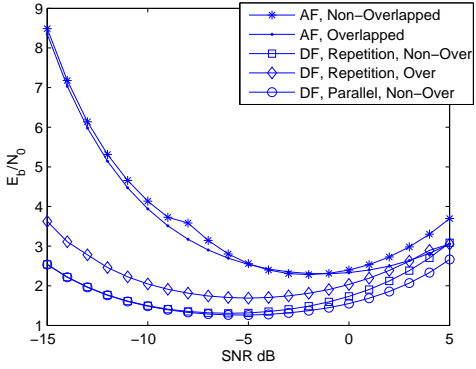}
\caption{Normalized bit energies $\frac{E_{b}}{N_{0}}$ vs. SNR for
the lowpass fading model when noncausal Wiener filter is employed.
$\sigma_{sd}^{2}=1,\sigma_{sr}^{2}=4$ and $\sigma_{rd}^{2}=4$.}
\label{fig:fig5}
\end{center}
\end{figure}
\begin{figure}
\begin{center}
\includegraphics[width = \figsize\textwidth]{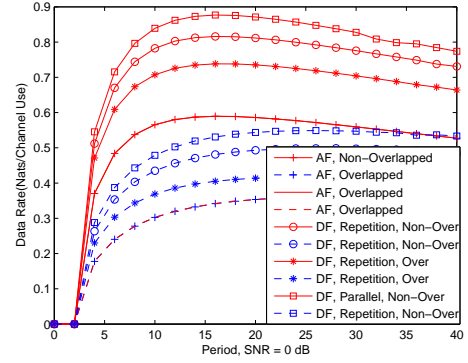}
\caption{The optimal achievable rates vs. training period $M$ for
the Gauss-Markov fading model. Single-pilot MMSE estimation is
employed. The dashed lines are obtained when
$\sigma_{sd}^{2}=1,\sigma_{sr}^{2}=4$ and $\sigma_{rd}^{2}=4$ and
solid lines are obtained when $\sigma_{sd}^{2}=1,\sigma_{sr}^{2}=16$
and $\sigma_{rd}^{2}=16$  } \label{fig:fig6}
\end{center}
\end{figure}
\begin{figure}
\begin{center}
\includegraphics[width = \figsize\textwidth]{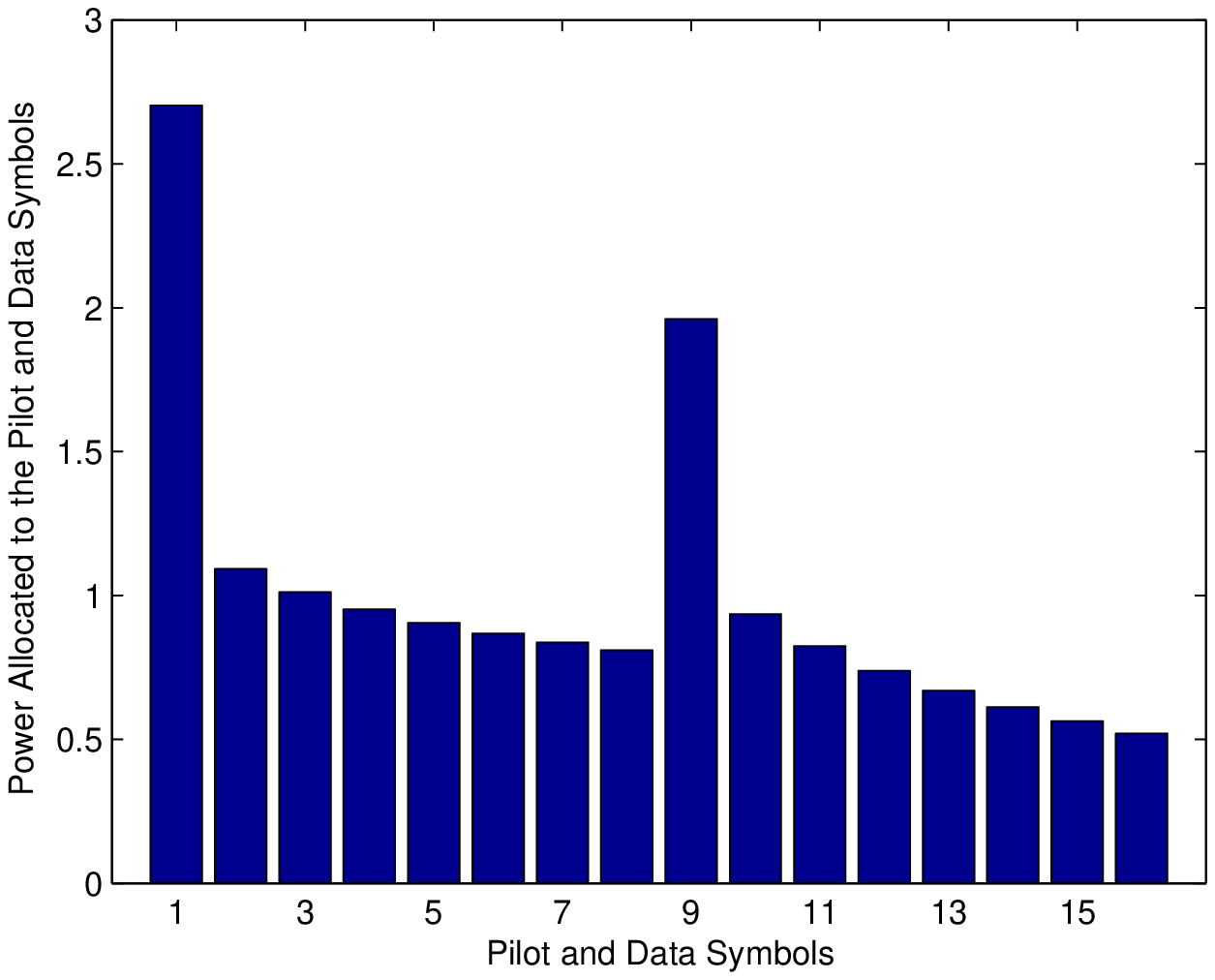}
\caption{Optimal power distribution among the pilot and data symbols
when $\sigma_{sd}^{2}=1,\sigma_{sr}^{2}=16$,$\sigma_{rd}^{2}=16$,
SNR=0dB. Fading is a Gauss-Markov process with $\alpha=0.99$.
Single-pilot MMSE estimation is employed. The optimal period is
$M=16$. Note that the first 8 symbols belong to the source and the
last 8 bars belong to the relay.} \label{fig:fig7}
\end{center}
\end{figure}
\begin{figure}
\begin{center}
\includegraphics[width = \figsize\textwidth]{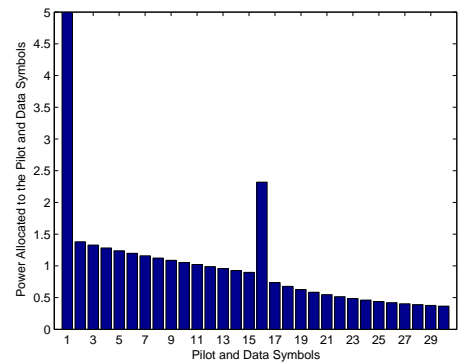}
\caption{Optimal power distribution among the pilot and data symbols
when $\sigma_{sd}^{2}=1,\sigma_{sr}^{2}=4$,$\sigma_{rd}^{2}=4$,
SNR=0dB. Fading is a Gauss-Markov process with $\alpha=0.99$.
Single-pilot MMSE estimation is employed. The optimal period is
$M=30$. Note that the first 15 symbols belong to the source and the
last 15 bars belong to the relay} \label{fig:fig8}
\end{center}
\end{figure}
In Fig. \ref{fig:fig2} and Fig. \ref{fig:fig3}, we plot the optimal
data rates when we estimate the lowpass fading process using a
noncausal Wiener filter. The channel variances are
$\sigma_{sd}^{2}=1,\sigma_{sr}^{2}=16,$ $\sigma_{rd}^{2}=16$, and
$\sigma_{sd}^{2}=1, \sigma_{sr}^{2}=4$ and $\sigma_{rd}^{2}=4$,
respectively. Conclusions similar to that given for Fig.
\ref{fig:fig1} are drawn again.
In Figs. \ref{fig:fig4} and \ref{fig:fig5}, the bit energy
normalized by the noise variance, $\frac{E_b}{N_0}$, is plotted as a
function of SNR. In all cases, we observe that minimum bit energy is
achieved at a nonzero SNR value. If SNR is further decreased, higher
bit energy values are required and hence, operation at these very
low SNRs should be avoided.

In Figure \ref{fig:fig6}, we plot the optimal data rates as a
function of the training period, $M$, when SNR=0 dB for different
relaying schemes and different channel variances.
Single-pilot-symbol estimation is employed. 
Since a relatively low SNR value is considered, AF non-overlapped
and AF overlapped schemes provide lowest rates. The highest
performance is obtained when DF parallel non-overlapped scheme is
used.

\begin{figure}
\begin{center}
\includegraphics[width = \figsize\textwidth]{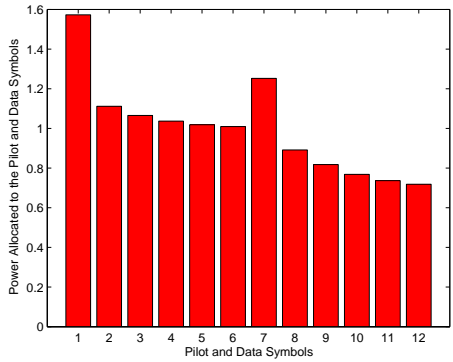}
\caption{Optimal power distribution among the pilot and data symbols
when $\sigma_{sd}^{2}=1,\sigma_{sr}^{2}=16$,$\sigma_{rd}^{2}=16$,
SNR=0dB. Fading is a Gauss-Markov process with $\alpha=0.99$. Wiener
filter is employed. The optimal period is $M=12$. Note that the
first 6 symbols belong to the source and the last 6 bars belong to
the relay} \label{fig:fig9}
\end{center}
\end{figure}
\begin{figure}
\begin{center}
\includegraphics[width = \figsize\textwidth]{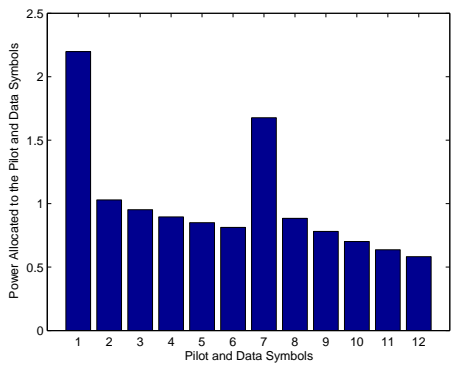}
\caption{Optimal power distribution among the pilot and data symbols
when
$\sigma_{sd}^{2}=1,\sigma_{sr}^{2}=16$,$\sigma_{rd}^{2}=16$,SNR=0dB.
Fading is a Gauss-markov process with $\alpha=0.99$. Single-pilot
MMSE estimation is employed. The optimal period $M=12$. Note that
the first 6 symbols belong to the source and the last 6 bars belong
to the relay} \label{fig:fig10}
\end{center}
\end{figure}

In Figs. \ref{fig:fig7} and \ref{fig:fig8}, power allocated to the
pilot and data symbols is plotted when Gauss-Markov channel is
considered and AF non-overlapped scheme is employed. The first half
of the bars shows the power allocated to the source symbols and the
rest shows the power allocated to the relay symbols. The first bar
of the each group gives the power of the pilot symbols. Note that
these power distributions are obtained when the period is at its
optimal value when SNR=0 dB. The optimal periods are 16 and 30 when
$\sigma_{sd}^{2}=1, \sigma_{sr}^{2}=16$, $\sigma_{rd}^{2}=16$, and
$\sigma_{sd}^{2}=1, \sigma_{sr}^{2}=4$, $\sigma_{rd}^{2}=4$,
respectively. In Figure \ref{fig:fig9}, the optimal power
distribution is displayed when noncausal Wiener filter is used for
estimation at SNR = 0 dB. Note that the optimal period is 12. In
Figure \ref{fig:fig10}, we plot the power distribution when
single-pilot estimation is performed at the optimal period 12 of
Wiener Filter estimation. It is observed that more power is given to
the pilot symbols when single-pilot-symbol estimation is used.
However, when we employ noncausal Wiener filter, the power allocated
to the pilot symbols is decreased thereby increasing the data rate
by giving more power to the data symbols.

\section{Conclusion}

We have studied transmission over imperfectly-known relay channels.
The channels are learned using single-pilot MMSE estimation or
noncausal Wiener filter. We have obtained achievable rate
expressions for both AF and DF relaying schemes. Subsequently, we
have jointly optimized the training period and power, and data power
levels in Gauss-Markov and lowpass fading models. We have compared
the performances of different relaying techniques at different SNR
values and different channel variances.


\end{document}